# EXPLORING HYDROGEN PRODUCTION FOR SELF-ENERGY GENERATION IN ELECTROREMEDIATION: A PROOF OF CONCEPT

*C. Magro[1*#], J. Almeida[1*], J.M. Paz-Garcia[2], E.P. Mateus[1] & A.B. Ribeiro[1#]*

[1]CENSE, Department of Sciences and Environmental Engineering, NOVA School of Science and Technology, NOVA University Lisbon, Caparica Campus, 2829-516 Caparica, Portugal
[2]Department of Chemical Engineering, Faculty of Sciences, University of Malaga, Teatinos Campus, 29010 Málaga, Spain

#corresponding authors: Cátia Magro c.magro@campus.fct.unl.pt; Alexandra B. Ribeiro abr@fct.unl.pt

*The authors had an equal contribution

## Highlights

- Self-produced $H_2$ from electrodialytic treatment of environmental matrices collected
- Collected $H_2$ average purity (% mol/mol) of ≈ 98%
- A fuel cell used to produce electricity from the self-produced $H_2$ (~1 V)
- Experimental self-generated energy promotes savings on electroremediation (≈ 7%)

## Abstract

Electrodialytic technologies are clean-up processes based on the application of a low-level electrical current to produce electrolysis reactions and the consequent electrochemically-induced transport of contaminants. These treatments inherently produce electrolytic hydrogen, an energy carrier, at the cathode compartment, in addition to other cathode reactions. However, exploring this by-product for self-energy generation in electroremediation has never been researched. In this work we present the study of hydrogen production during the electrodialytic treatment of three different environmental matrices (briny water, effluent and mine tailings), at two current intensities (50 and 100 mA). In all cases, hydrogen gas was produced with purities between 73% to 98%, decreasing the electrical costs of the electrodialytic treatment up to ≈ 7%. A proton-exchange membrane fuel cell was used to evaluate the possibility to generate electrical energy from the hydrogen production at the cathode, showing a stable output (~1 V) and demonstrating the proof of concept of the process.

## 1. Introduction

Global energy demands from an increasing human population is a major concern for the planet sustainability. Extensive research and technology development have been focused on renewable energy sources and other strategies to reduce $CO_2$ emissions [1]. Fuel cell technology, which can efficiently generate electricity using hydrogen as fuel, has attracted widespread attention in recent years [2]. The proton-exchange membrane fuel cells success depends on their ability to obtain optimal fuel to electricity conversion with a high current density, as well as the sustainable and economical production of the fuel [3]. Pure hydrogen gas is scarce in Earth's atmosphere. However, it can be produced from different primary-energy sources. For instance, it can be generated from fossil fuels through steam reforming, partial oxidation or gasification and from renewable sources through biomass gasification and water electrolysis [4,5]. Generation of $H_2$ via water electrolysis is still limited by the high cost, namely ≈ 3.8 times more expensive than gasification, and ≈ 5 times more expensive than from methane steam reforming [6]. Hence, steam reforming, which combines high-temperature steam with natural gas, currently accounts for the majority of the $H_2$ produced. Hydrogen production via water electrolysis is currently only applied in combination with renewable energy sources, like solar or wind, and used as an energy storage system.

Electro-based technologies, such as electrokinetic and electrodialytic processes, have been the focus of vast environmental remediation research over the last three decades [7,8], both *in-situ* [9,10], and *ex-situ* [11, 12, 13]). Despite such research efforts, the technology readiness level (TRL) for many of those technologies remains very low; although most are considered promising, many are far from being introduced as efficient processes into the market. Important barriers need to be overcome to reach high TRLs [14]. Operational energy costs have to be considered and, are related not only to the electrolysis reactions but mainly with the stirring, the ohmic losses and the energy required for the transport of charge through the porous matrix. In fact, the distance between electrodes (cell size) plays a crucial role in the energy costs of the specific-energy required for the target contaminants removal [7,15]. To the best of our knowledge, there has been minimal research conducted related to the reuse of the elemental gases produced in the electrolysis reactions during electrochemically-induced treatments. The drawbacks found in the current literature are associated to the reactors' design. Most electrokinetic and electrodialytic (ED) setups are designed to allow for the produced gases to flow freely into the atmosphere, while aiming to reduce pressure-related transport mechanisms. Thus, a gas collection strategy during the treatment is not included in the system, causing gas losses to the atmosphere. As a novel feature, the $H_2$ produced during the treatment at the cathode compartment may be used as fuel in a proton-exchange membrane fuel cells to produce electrical energy and reduce the energy costs of electroremediation. Additionally, as an energy carrier, $H_2$ can be used to accumulate energy during the electric power demand valleys, and to generate electric power during the peaks.

Therefore, a reservoir can be integrated into the ED system where it can recover and use the $H_2$ produced for different purposes.

This work evidences the possibility of using the $H_2$ produced during electrochemically-induced remediation of three different environmental matrices: (1) moderately-salted water - briny water, (2) secondary effluent from a wastewater treatment plant, and (3) mine tailings. Our proof of concept demonstrates that the $H_2$ captured and reused from these ED treatments is feasible. Herein, a three-compartment ED set-up was used to minimize the interactions of the sample and the contaminants with the electrolysis reactions (Figure 1).

## 2. Theory

### 2.1. Proton-exchange fuel cell

A fuel cell is an electrochemical device that converts the chemical energy from a fuel into electricity through the reaction of the fuel with oxygen or another oxidizing agent. For example, a proton-exchange membrane fuel cell (PEMFC) combines $H_2$ and $O_2$ to produce electricity and heat without emitting other products which are different from the water formed in the reaction eq. (1) [16]:

$$H_2 + \frac{1}{2}O_2 \rightarrow H_2O \tag{1}$$

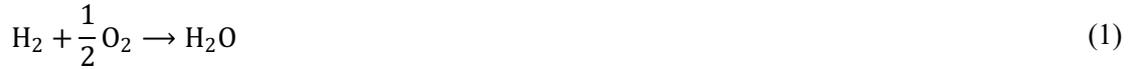

A fuel cell, unlike a battery, produces electricity as long as fuel is supplied, never losing its charge. The pollution-free production of energy and high power density makes the fuel cell technology a viable approach for future energy industries [2]. Fuel cells show high energy conversion efficiency, up to 60%, higher than traditional internal combustion engines [17]. This efficiency can increase up to 80% with heat-recovery systems [18].

### 2.2 Electro-based technologies

Electrokinetic and ED strategies are commonly applied to remove organic [19] and/or inorganic contaminants from soils or other porous matrices, such as sewage sludge, fly ash or construction materials [20, 21, 22]. The electrochemically-induced transport is based on the application of a low level direct current which promotes electrolysis reactions at the electrodes [20,23,24], involving in most cases water oxidation at the anode, eq. (2), and water reduction at the cathode, eq. (3):

$$O_2(g) + 4H^+ + 4e^- \rightarrow 2H_2O(l); E^o_{anode}(25\ °C) = 1.23\ V \tag{2}$$

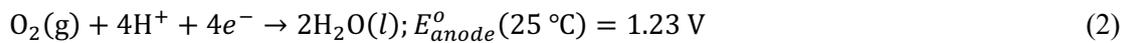

$$H_2(g) + 2OH^- \rightarrow 2H_2O(l) + 2e^-; E^o_{cathode}(25°C) = 0.83\ V \tag{3}$$

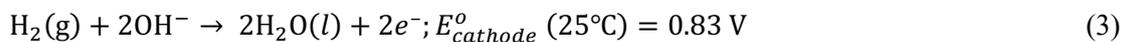

Competing redox reactions may occur as, for example, the production of chlorine at the anode in systems with high chloride contents [25], eq. (3):

$$Cl_2(g) + 2e^- \rightarrow 2Cl^- \; ; E^o_{anode} (25°C) = 1.36 \text{ V} \qquad (4)$$

or the deposition of metals (Me) at the cathode, eq. (3):

$$Me^{2+} + 2e^- \rightarrow Me^o \qquad (5)$$

The electrochemical-induced transport of chemical species takes place by three main transport mechanisms: electromigration, electroosmosis, and electrophoresis. Diffusion and advection may also play an important role [20]. In the case of the ED process, electrodialysis also occurs, as ion-exchange membranes are used to separate the matrix from the electrode compartments (aiming to control the pH conditions of the electrolytes and the treated matrix while improving the selectivity on the contaminant removal [20]). Over the years, different electro-based remediation set-ups have been proposed, where the configuration of the sample and the electrode compartments have been modified depending on the nature of the contaminant and matrix [26,27].

## 3. Materials and Methods

### 3.1. Materials

The briny water solution was prepared with NaCl (PA grade, Merck, Germany) and tap water (Almada, Portugal). Effluent, the liquid fraction that results from wastewater treatments, was collected in the secondary clarifier at a wastewater treatment plant (Lisbon, Portugal). Mine tailings were collected at Panasqueira mine (Covilhã, Portugal, 40°10'11.0604"N, 7°45'23.8752"W). Panasqueira mine produces around 900 t $WO_3$/year [28] and the pond where the residues are deposited is an open air impoundment that contains rejected ore concentrates with high metal levels [29]. The matrix used for this study is a rejected fraction from the sludge circuit, that is directly pumped to the Panasqueira dam.

### 3.2. Experimental set-up

The ED cell set-up was a 3 compartment acryl XT cell [30] (RIAS A/S, Roskilde, Denmark), as represented in Figure 1. The internal diameter was 8 cm and the central and electrolyte compartments length were 5 cm. The two electrode compartments were separated from the central section by an anion exchange membrane, AR204SZRA, MKIII, Blank (Ionics, USA) and a cation exchange membrane, CR67, MKIII, Blank (Ionics, USA). The electrodes were made of Ti/MMO Permaskand wire: $Ø = 3$ mm, $L = 50$ mm (Grønvold & Karnov A/S, Denmark). Ti/MMO anodes are used to degrade organic contaminants in wastewaters, and Ti/MMO is also applied as cathode to reduce chlorinated and nitro compounds in groundwater [31]. A power supply E3612A (Hewlett Packard, Palo Alto, USA) was used to maintain a constant current in the ED cell.

For briny water and effluent experiments, 250 mL of liquid sample was added to the central cell compartment. For the mine tailings experiments, suspensions were prepared at a liquid/solid (L/S) ratio of 9, by mixing 22.2 g of solid mine tailings within 200 mL of briny water. The anolyte and catholyte compartments were set with 250 mL of 0.01M $NaNO_3$.

Twelve ED experiments were carried out in duplicate according to the conditions presented in Table 1. In experiments 1-6 the gas produced at the cathode, rich in $H_2$, was collected in a 30 mL storage cylinder (Horizon Fuel Cell Technologies, USA) (experimental scheme at supplementary data B.1), where the volume was verified every 10 minutes. In experiments 7-9, the cathode compartment exhaust was directly connected to the tedlar sample bag, single polypropylene fitting with 500 mL of capacity (SKC, USA), for purity analysis. In experiments 10-12, the cathode compartment exhaust was directly connected to the PEMFC. In all cases, the ED cell voltage and the fuel cell open circuit voltage were registered every 10 minutes. The fuel cell open circuit voltage was measured in order to validate the $H_2$ catchment and conversion into power needs.

The single PEMFC (Horizon Fuel Cell Technologies, USA) was used (32 × 32 × 10 mm), with a nominal voltage of ≈ 1 V. The PEMFC has a cathodic plate, designed as a part of the cell's membrane electrode assembly that collects $O_2$ directly from the air by natural convection. PEMFC voltage and resistance were measured and monitored by a multimeter KT1000H (KIOTTO, Portugal).

### 3.3 Methods

pH and conductivity were measured at the beginning and at the end of all ED experiments, both in central and electrode compartments. Briny water, effluent and mine tailings pH were measured with a Radiometer pH-electrode EDGE (HANNA, USA), and conductivity was measured in a Radiometer Analytic LAQUA twin (HORIBA Ltd., Japan). The mine tailings pH and conductivity measurements are referred to the liquid phase resulting from the suspension from a liquid component (either deionized $H_2O$ or briny water), with a L/S ratio of 9.

Total concentrations of As, Ca, Cu, K, Mg, Na, P, S, Sn, W, and Zn were determined by Inductively Coupled Plasma with Optical Emission Spectrometry (ICP-OES) (HORIBA Jobin-Yvon Ultima, Japan), equipped with generator RF (40.68 MHz), monochromator Czerny-Turner with 1.00 m (sequential), automatic sampler AS500 and dispositive CMA-Concomitant Metals Analyser. $Cl^-$ and $SO_4^{2-}$ were analyzed by Ion Chromatography (IC) (DIONEX ICS-3000, USA), equipped with conductivity detector. To quantify the elements in the solid matrix, an acid extraction was carried out mixing 0.5 g of mine tailings, 3 mL of HCl (37%) and 9 mL of $HNO_3$ (65%) and, placed on a shaking table for 48 hours at 125 rpm. For the IC analysis of the mine tailings ($Cl^-$ and $SO_4^{2-}$ content), microwave assisted acid extraction was carried out according to EPA method 3051 A: 0.5 g of mine tailings were placed in a vessel with 3 mL of HCl (37%), and 9 mL of $HNO_3$ (65%) and placed in a microwave Ethos (Milestone S.r.l, Bergamo, Italy). At the end, all the samples were diluted in deionized water (1:25), filtered by vacuum using 0.45 μm MFV3 glass microfibre filters (Filter lab, Barcelona, Spain) and analyzed by ICP-OES and IC.

The $H_2$ purity percentage was determined by Gas Chromatography Thermal Conductivity Detector (GC-TCD) using a Trace GC Ultra (Thermo Electron Corporation, USA), with a Carboxen 1010 plot column (length: 30 m, diameter: 0,32 mm). The analytical run was performed in an isothermal mode at 35 ºC for 50 minutes. A gastight syringe (vici precision sampling, Baton Rouge, Lousiana, USA) was used to inject a volume of 250 μL (injector at 200 ºC), detector/transfer line at 120 ºC. To calculate the $H_2$ purity two methods were used: (1) internal linear calibration and response factor ($H_2$ peak area/response factor), where the standard deviation is related to error of these methods, by comparing with the injection of 100% $H_2$; (2) molar proportions (mol/mol), were determined assuming air as impurity in the $H_2$. Thus, the $H_2$ was calculated considering the number of $H_2$ mol in 100 mol of air ($H_2$ + air gases).

All sample analysis was carried out in duplicate. The data from the experiments were analyzed by the software Origin Pro 8.5 and the statistical data obtained by the GraphPad Prism version 7.0e. Statistically significant differences among samples for 95% level of significance were calculated through ANOVA tests.

## 4. Results and discussion

### 4.1. Matrix characterization

Matrices selection is an important step as the matrices' characteristics will affect $H_2$ purity and further energy generation. PEMFC, despite the robustness and stability, may be sensitive to contaminants in the fuel [32]. The briny water was chosen as a working system blank, where NaCl was added to emulate the effluent without interferences. Briny water, with sufficient ionic conductivity to maintain the current applied for the remediation period chosen (1 and 2h), is typically used in systems for the electrolytic production of $H_2$. The effluent and mine tailings are, individually, matrices with high potential to be reused as raw materials in several sectors. For example, ED treated effluent has recently been tested for cement based construction materials [33]. Mine tailings are an example of solid matrix that can be successfully treated via ED as stirred suspensions mixtures [34]. In these cases, $H_2$ production and exploitation is highly attractive, since it can allow the decrease of energy costs in an industrial scale application.

Table 2 presents the initial characterization of the three studied matrices. The matrices had enough initial conductivity to allow current passage and facilitate the electrolysis reactions to occur at the imposed rate. The initial pH of the studied matrices was in the range 4.6 - 7.7. Mine tailings were slightly acidic (pH ≈ 4.57), presenting a high concentration of arsenic (218.57 ± 132.31 mg As/kg), and significant amounts of other metals (76.82 ± 39.30 mg Cu/kg, 1.95 ± 0.53 mg Sn/kg and 5.34 ± 1.42 mg W/kg), as well as a high sulfur content (240.9 ± 4.6 mg/kg).

The ED process was applied to briny water, effluent and mine tailings suspension in briny water. In all cases, the pH at the anode compartment decreased to ~ 2 and the pH at the cathode compartment increased until ~ 12. The pH at the central compartment also decreased in all cases. This phenomenon was expected as anion exchange membranes are known to have limited perm-selectivity, which means that only protons ($H^+$) are able to cross this membrane. The acidification of the central compartment was more noticeable in the experiments at 100 mA (Figure 2). The acidification phenomena proved to be influenced by the current intensity. The smallest variation in the potential applied in the ED cell will increase the $H^+$ production in the media. Thus, not only the current intensity should be considered to analyze the results, but also the error associated to the power supplier (± 3 mA).

Final concentrations of the major concerned elements in the liquid matrices ($Cl^-$ and $SO_4^{2-}$) and in the mine tailings suspension (As, Cu, Sn and W) are presented in Tables 3 and 4, respectively.

In general, the target contaminants removal from the matrices was higher in the experiments operated at 100 mA, for the same amount of circulated charge. When ED treatment is performed at higher current intensities, the electromigration transport is predominant over diffusion or other transportation phenomena. The amount of salts amount is extremely heterogeneous in the effluent. The season periods of the sampling procedure affected chemical and physical properties of the samples collected due to the fluctuations in weather conditions and also the wastewater treatment plant process efficiency.

The slightly extraction of Sn and W from mine tailings may be related to specific chemical limitations, such as desorption or dissolution mechanisms. For example, W complexes are quickly decomposed and stabilized by high concentrations of chloride ions (MT: 5.6 ± 2.3 mg/kg; BW: 499.3 ± 8.1 mg/kg), where the most common product of the decomposition is $[W_2Cl_9]^{3-}$. Alternatively, the adsorption of sulfate ions on metallic W surface results in the electronic structure modification. The $O_2$ reduction reaction is blocked during the ED process, which is already low in cationic dissolution in electrochemical processes [35].

### 4.2. Hydrogen generation and use

Figure 3 shows the volume of collected gas produced at the cathode compartment during experiments 1–6. As the volume of the gas deposit was 30 mL, it was filled before the ED process ended. Experiments at 50 mA showed matrix related differences in the gas flow rate, reaching the 30 mL of $H_2$ production at different rates. There are no statistically significant differences for the flow rates obtained at 100 mA (Table E at supplementary data). However, the flow rates of $H_2$ are significantly different ($p < 0.001$) in the experiments at 50 mA.

The gas generated at the cathode compartment, in the experimental setup 3, 4 and 6 during 6 h, was captured and analyzed via GC-TCD, in a tedlar sample bag. Table 5 presents the purity of the captured gas for these experiments carried out at 100 mA.

The average $H_2$ purity (% w/w) of the produced gas was ≈ 73% (w/w), where the highest value observed was in the ED experiment applied to the effluent, that produced a gas with 90.4 ± 0.3% of $H_2$. Thus, only the $H_2$ purity in the effluent experiment has a statistically significant difference ($p < 0.05$) compared to briny water and mine tailings suspension, while briny water and mine tailings suspension does not have a statistically significant difference between each other (Table 5). In order to validate and make a comparison with the purity results obtained in % w/w, another approach to determine the $H_2$ purity was carried out. Thus, the calculation of the gas was also performed considering the molar fraction of the $H_2$ in the air gases. Comparing the two methods, the % mol/mol of $H_2$ of the produced gas was 19% higher, in average, comparing to the % w/w (Table 5). In the % mol/mol analysis, the purity of the $H_2$ was higher than 97% in all samples, whereas in the % w/w the $H_2$ purity oscillated between 72% and 90%. The mass and the molar compositions are different, and it is expected a higher molar purity, as long as the other components in the gaseous phase are heavier (e.g. $N_2$, $O_2$).

As mentioned before, the flow rate production for $H_2$ in the ED treatments can be directly related to the current intensity and the matrix. Assuming the $H_2$ captured at 1 atm and 25 °C, a total of ≈ 45.6 mL of pure dry $H_2$ would be theoretically obtained, at a rate of 0.76 mL/min in the experiments at 100 mA. The volume of collected gas at the cathode, shown in Figure 3 (Table B.2 at supplementary data), was clearly higher than the expected during the first few minutes of treatment, with a change in the production rate after the first 10 to 20 minutes. This may indicate that competing cathode reactions took place. For example, a possible electrode reaction that could have taken place is the formation of $NO_2$ (g) from the reduction of the nitrates ($E_0 = 0.803$ V), shown in eq. (6).

$$NO_2(g) + H_2O \rightarrow NO_3^-(aq) + 2H^+ + e^- \quad ; E^o_{cathode}(25°C) = 0.803 \text{ V} \tag{6}$$

This reaction produces 1 mol of $NO_2$ (g) per circulated electron, twice as much as the $H_2$ reaction and would increase the pH of the cathode compartment twice as fast. However, the pH changes at the cathode and the decrease of nitrate concentration in time will promote that the water reduction electrolysis reaction become predominant after the first several minutes of the reaction. $NO_2$ (g) might be in a solution in the form of $N_2O_4$ below 21 °C.

### 4.3. Electrical energy requirements in the ED experiments

Among the ED experiments with $H_2$ capture (experiments 1–6 in Table 1), the ED cell voltage decreased overtime. The rapid decrease of the cell voltage was more evident in the experiments at 100 mA (Figure 4).

According to the Ohm's law, if the current intensity *(I)* is constant, the voltage (V) and the resistivity (R, or conductivity) are strongly related, eq. (7):

$$V = RI \text{ (V)} \tag{7}$$

Where *I* is the current intensity, V the voltage and, R is the resistance.

The decrease of the ED cell voltage is related to a conductivity increase in the electrode compartments. The initial conductivity of the electrolyte was moderately low, $0.90 \pm 0.06$ mS/cm, and it increased to values between 3 - 4 mS/cm at the anode, and 2 - 3 mS/cm at the cathode, at the end of all experiments, which is consistent with the pH changes observed. In the central compartment, the conductivity changed depending on the matrix (Figure 5). Due to the highly heterogeneous environmental samples under study, high standard deviations in the ED experiments' behavior were observed. Nevertheless, in general, it stayed with values that assure conductivity. However, the experiment MTBW at 100 mA had a voltage increase after half an hour and a decrease on the conductivity at the end of the experiment (Figure 5). The treatment at 100 mA produced the reduction of ions (related to the $H^+$ migration through the anion exchange membrane) in the central compartment, and an energy efficiency decrease of the process due to an increase of the ohmic losses. In the experiments at 50 mA, the longer running times allowed diffusion and dissolution/desorption kinetics to replete the ion content in the central cell compartment. Effluent and mine tailings are matrices with high dependency on the sampling and weather conditions (effluent high standard deviations in the salts content, in line with the conductivity values). Mine tailings are a heterogeneous matrix implying different amounts of metals and salts content, with different variations in the charged species during the sampling.

As the ED experiments are carried out at a constant current, the electrical energy required during the process, accounting only for the energy applied by the DC power source, can be calculated from eq. (8):

$$E = I \int_{t_0}^{t} V_{cell}(t)\, dt \tag{8}$$

Using a numerical chained-trapezoidal integration, the estimated electrical energy is presented in Figure 6. As expected, the electrical energy required increases as the electrical current increases, for the same amount of circulated charge. These results are consistent with the conductivity of those matrices. According to the results presented in the Figure 6, the experiments carried out at 100 mA required $\sim 8.7 \pm 0.8$ kJ of electrical energy, while the experiments carried out at 50 mA required an average of $\sim 4.8 \pm 0.7$ kJ.

### 4.4 Energy Savings

Considering a faradaic efficiency of 100% and no competition to the water reduction as the cathode electrolysis reaction, ~ 1.86 mmol of $H_2$ would be produced in experiments 1-6 at the end of the ED experiment, either at 50 mA in 2 h or at 100 mA in 1 h. Using a fuel cell at low temperatures and considering the higher heat of combustion of $H_2$ as 141.8 MJ kg$^{-1}$, a total of 0.53 kJ may have been recovered from the process. The efficiency of the chemical energy conversion to electrical energy at the fuel cell depends on the quality of the $H_2$ gas produced, where a maximum of 98% (mol/mol) hydrogen average purity was achieved during the ED treatments. Considering that current fuel cell has energy conversion efficiencies of around 60%, a total of 0.32 kJ could have been transformed into electrical energy. This translates to ~ 3.1% of the electrical energy required in experiment 4 (EF 100 mA, the worst case), and ~ 6.9% of the electrical energy required in experiment 5 (MTBW 50 mA, the best case). Reducing the electrical energy dependency, mainly due to stirring needs (for solid fine matrices), transport of charge through the porous matrix, and to feed a power supply for the electrode's reactions, may also have impact in the total variable costs of the overall ED process (~7%).

As expected, the higher the current intensity used in the ED treatment, the higher the electrical energy requirements for the same amount of circulated charge. In the experiments presented, those carried out at 50 mA during 2 h required almost half of the electrical energy to produce the electrolysis reactions than those carried out at 100 mA during 1 h. On the other hand, higher applied currents obtain better removal efficiencies, provided that there are no phase-transfer kinetic limitations, such as dissolution or desorption processes.

In general, one of the main advantages of the ED treatment applied to liquid matrices compared to ED treatments applied solid porous matrices is that the energy requirements for the electrolysis reactions are considerably smaller, due to the higher conductivity of the matrices. Thus, in solid porous matrices, the ED treatment would reach higher voltage gradients due to the lower conductivity and will require long-lasting treatments. The ED treatments presented show requirements of electrical power for the electrolysis in the range of 1 to 5 W. However, ED treatments carried out for the mine tailings suspension required additional energy for the stirring system at the sample central cell compartment. A 10 W stirrer system was used, meaning that the electrical energy required for the stirring system could be up to ten-fold compared to the energy required for the electrolysis reactions. For the purpose of energy optimization in ED treatments applied to suspensions, the reduction of stirring costs is critical. In this sense, PEMFC could be an important factor in ED energy savings by reducing the operation costs by powering a low-energy stirring system.

To evaluate the possibility of reducing energy requirements from the ED treatments by the *in-operando* production of electricity from $H_2$ gas formed at the cathode, a PEMFC was connected directly to the exhaust pipe of the cathode compartment in experiments 7–9, corresponding to the same matrices under ED treatment at 100 mA during 1 h.

The initial open circuit voltage of the PEMFC was, in all cases, near 1.4 V, and it decreased and stabilized at a value of ~ 1 V (Figure 7), as expected for a single PEMFC [32]. To obtain higher voltages, a stack of FCs connected in series could be used. Comparing the voltage generated by PEMFC in the different cases, there are no statistically significant differences ($p < 0.05$) (see Table C at supplementary data). The data supported the statement that the production of electricity by a PEMFC is independent of environmental matrices used in the ED treatments presented, despite the fluctuation observed for the case of effluent.

The data presented in this study indicates that the gas produced at the cathode has a purity between 72.4% and 99.3%. Therefore, in a field scale ED treatment, the produced $H_2$ could also be stored and sold for transportation or other industrial sectors. The production costs of $H_2$ via electrolysis vary around 8-11 €/kg, which is higher than that obtained via steam methane reforming using natural gas or biogas [36,37]. The $H_2$ produced during ED treatments, that until now has been an unexploited byproduct, may be an alternative source of $H_2$ for transportation or energy storage.

## 5. Conclusions

In this study it is proved the possibility to produce $H_2$ with average purities between 73% and 98% from electrodialytic treatments and used to generate electrical energy with a proton-exchange membrane fuel cell. This estimation was performed according to the $H_2$ formed at the cathode electrolysis reaction, which can reduce the energy costs associated to the electrodialytic treatments, as well as any other remediation treatment based on electrochemically-induced transport.

The efficiency of chemical to electrical energy conversion at the fuel cell would depend on the quality of the $H_2$ gas produced. Thus, specific studies on possible competitive cathode reactions are needed depending on the system. However, in the matrices tested in this study, $H_2$ gas was produced in all cases. This suggests that $H_2$ purity seems to be more affected by external factors (experimental errors, nitrate reduction to $NO_2$ or temperature increase) than the matrices composition. Flowing the produced $H_2$ gas through a single proton-exchange membrane fuel cell, resulted in a stable open circuit voltage (~1V), that demonstrated the potential to recover energy from the $H_2$ byproduct, that otherwise would be released to the atmosphere and lost.

This research shows there are new possibilities for energy saving and $H_2$ production for different purposes in electrodialytic treatment, leading to an increase in the sustainability and applicability of the electro-remediation, decontamination or degradation contaminants´ processes.

**Glossary**

   ED              Electrodialytic

| | |
|---|---|
| GC-TCD | Gas Chromatography with Thermal Conductivity Detector |
| IC | Ion Chromatography |
| ICP-OES | Inductively Coupled Plasma with Optical Emission Spectrometry |
| PEMFC | Proton-exchange membrane fuel cell |

**Acknowledgments**

This work has received fundings from the European Union's Horizon 2020 research and innovation programme under Grant Agreement No. 776811, under the Marie Skłodowska-Curie grant agreement No. 778045, and from Portuguese funds from FCT/MCTES through grants UID/AMB/04085/2019 and PTDC/FIS-NAN/0909/2014. C. Magro and J. Almeida acknowledge Fundação para a Ciência e a Tecnologia for their PhD fellowships SFRH/BD/114674/2016 and PD\BD\135170\2017, respectively. The authors acknowledge Carla Rodrigues and Nuno Costa from REQUIMTE for the ICP and IC analysis, and Ricardo Chagas from LAQV- REQUIMTE.

**References**

[1] Hart PS, Feldman L. Would it be better to not talk about climate change? The impact of climate change and air pollution frames on support for regulating power plant emissions. J Environ Psychol 2018:1–33. doi:10.1016/j.jenvp.2018.08.013.

[2] Pei P, Jia X, Xu H, Li P, Wu Z, Li Y, et al. The recovery mechanism of proton exchange membrane fuel cell in micro-current operation. Appl Energy 2018;226:1–9. doi:10.1016/j.apenergy.2018.05.100.

[3] Majlan EH, Rohendi D, Daud WRW, Husaini T, Haque MA. Electrode for proton exchange membrane fuel cells: A review. Renew Sustain Energy Rev 2018;89:117–34. doi:10.1016/j.rser.2018.03.007.

[4] Abdalla AM, Hossain S, Nisfindy OB, Azad AT, Dawood M, Azad AK. Hydrogen production, storage, transportation and key challenges with applications: A review. Energy Convers Manag 2018;165:602–27. doi:10.1016/j.enconman.2018.03.088.

[5] Nikolaidis P, Poullikkas A. A comparative overview of hydrogen production processes. Renew Sustain Energy Rev 2017;67:597–611. doi:10.1016/j.rser.2016.09.044.

[6] Baykara SZ. Hydrogen: A brief overview on its sources, production and environmental impact. Int J Hydrogen Energy 2018;43:10605–14. doi:10.1016/j.ijhydene.2018.02.022.

[7] Alshawabkeh AN. Electrokinetic soil remediation: Challenges and opportunities. Sep Sci Technol 2009;44:2171–87. doi:10.1080/01496390902976681.

[8] Ribeiro AB, Mateus EP, Couto N, editors. Electrokinetic Across Disciplines and Continents. New Strategies for Sustainable Development, Springer International Publishing; 2016, 427 pp., doi:10.1007/978-3-319-20179-5.

[9] Gomes HI, Dias-Ferreira C, Ribeiro AB. Overview of in situ and ex situ remediation technologies for PCB-contaminated soils and sediments and obstacles for full-scale application. Sci Total Environ 2013;445–446:237–60. doi:10.1016/j.scitotenv.2012.11.098.

[10] Yukselen-Aksoy Y, Reddy KR. Effect of soil composition on electrokinetically enhanced persulfate oxidation of polychlorobiphenyls. Electrochim Acta 2012;86:164–9. doi:10.1007/s12520-016-0397-x.

[11] Ferreira C, Jensen P, Ottosen L, Ribeiro A. Removal of selected heavy metals from MSW fly ash by the electrodialytic process. Eng Geol 2005;77:339–47. doi:10.1016/j.enggeo.2004.07.024.

[12] Kirkelund GM, Magro C, Guedes P, Jensen PE, Ribeiro AB, Ottosen LM. Electrodialytic removal of heavy metals and chloride from municipal solid waste incineration fly ash and air pollution control residue in suspension - Test of a new two compartment experimental cell. Electrochim Acta 2015;181:73–81. doi:10.1016/j.electacta.2015.03.192.

[13] Ebbers B, Ottosen LM, Jensen PE. Electrodialytic treatment of municipal wastewater and sludge for the removal of heavy metals and recovery of phosphorus. Electrochim Acta 2015;181:90–9. doi:10.1016/j.electacta.2015.04.097.

[14] Lacasa E, Cotillas S, Saez C, Lobato J, Cañizares P, Rodrigo MA. Environmental applications of electrochemical technology. What is needed to enable full-scale applications? Curr Opin Electrochem 2019;16:149–56. doi:https://doi.org/10.1016/j.coelec.2019.07.002.

[15] Villen-Guzman M, Gomez-Lahoz C, Garcia-Herruzo F, Vereda-Alonso C, Paz-Garcia JM, Rodriguez-Maroto JM. Specific Energy Requirements in Electrokinetic Remediation. Transp Porous Media 2018;121:585–95. doi:10.1007/s11242-017-0965-2.

[16] Kone J-P, Zhang X, Yan Y, Hu G, Ahmadi G. CFD modeling and simulation of PEM fuel cell using OpenFOAM. Energy Procedia 2018;145:64–9. doi:10.1016/j.egypro.2018.04.011.

[17] Staffell I, Green R, Kendall K. Cost targets for domestic fuel cell CHP. J Power Sources 2008;181:339–49. doi:10.1016/j.jpowsour.2007.11.068.

[18] Ahmadi MH, Mohammadi A, Pourfayaz F, Mehrpooya M, Bidi M, Valero A, et al. Thermodynamic analysis and optimization of a waste heat recovery system for proton exchange membrane fuel cell using transcritical carbon dioxide cycle and cold energy of liquefied natural gas. J Nat Gas Sci Eng 2016;34:428–38. doi:10.1016/j.jngse.2016.07.014.

[19] Cameselle C, Gouveia S. Electrokinetic remediation for the removal of organic contaminants in soils. Curr Opin Electrochem 2018;11:41–7. doi:10.1016/j.coelec.2018.07.005.

[20] Ribeiro AB, Rodríguez-Maroto JM. Electroremediation of heavy metal-contaminated soils. Processes and applications. Cap. 18. In: Prasad MNV, Sajwan KS, Naidu R, editors. Trace Elements in the Environment: Biogeochemistry, Biotechnology and Bioremediation, Taylor & Francis, Florida, USA: CRC Press; 2006, p. 341–68.

[21] Magro CC, Guedes PR, Kirkelund GM, Jensen PE, Ottosen LM, Ribeiro AB. Incorporation of different fly ashes from MSWI as substitute for cement in mortar: An overview of the suitability of electrodialytic pre-treatment. In: Ribeiro AB, Mateus EP, Couto N, editors. Electrokinetic Across Disciplines and Continents. New Strategies for Sustainable Development, Springer International Publishing; 2016, p. 225–47. doi:10.1007/978-3-319-20179-5_12.

[22] Guedes P, Mateus EP, Couto N, Rodríguez Y, Ribeiro AB. Electrokinetic remediation of six emerging organic contaminants from soil. Chemosphere 2014;117:124–31. doi:10.1016/j.chemosphere.2014.06.017.

[23] Guedes P, Couto N, Ottosen LM, Ribeiro AB. Phosphorus recovery from sewage sludge ash through an electrodialytic process. Waste Manag 2014;34:886–92. doi:10.1016/j.wasman.2014.02.021.

[24] Villen-Guzman M, Paz-Garcia JM, Rodriguez-Maroto JM, Gomez-Lahoz C, Garcia-Herruzo F. Acid Enhanced electrokinetic remediation of a contaminated soil using constant current density: Strong vs. weak acid. Sep Sci Technol 2014;49:1461–8. doi:10.1080/01496395.2014.898306.

[25] Paz-García JM, Johannesson B, Ottosen LM, Alshawabkeh AN, Ribeiro AB, Rodríguez-Maroto JM. Modeling of electrokinetic desalination of bricks. Electrochim Acta 2012;86:213–22. doi:10.1016/j.electacta.2012.05.132.

[26] Hansen HK, Rojo A, Ottosen LM. Electrodialytic remediation of copper mine tailings. Procedia Eng., vol. 44, 2012, p. 2053–5. doi:10.1016/j.proeng.2012.09.042.

[27] Pedersen KB, Jensen PE, Ottosen LM, Barlindhaug J. Influence of electrode placement for mobilising and removing metals during electrodialytic remediation of metals from shooting range soil. Chemosphere 2018;210:683–91. doi:10.1016/j.chemosphere.2018.07.063.

[28] Lefebvre G, García R, Arragonés M, Moya M, Monge Q, Maund N. Report on balance between demand and supply of refractory metals in the EU. MSP-REFRAM 2016;D1.3:40.

[29] Ávila PF, Silva EF Da, Salgueiro AR, Farinha JA. Geochemistry and mineralogy of mill tailings impoundments from the panasqueira mine (Portugal): Implications for the surrounding environment. Mine Water Environ 2008;27:210–24. doi:10.1007/s10230-008-0046-4.

[30] Ribeiro AB, Mexia JT. A dynamic model for the electrokinetic removal of copper from a polluted soil. J Hazard Mater 1997;56:257–71. doi:10.1016/S0304-3894(97)00060-5.

[31] Yuan S, Gou N, Alshawabkeh AN, Gu AZ. Efficient degradation of contaminants of emerging concerns by a new electro-Fenton process with Ti/MMO cathode. Chemosphere 2013;93:2796–804. doi:10.1016/J. chemosphere.2013.09.051.

[32] Joshua OS, Ejura GJ, Essien VE, Olokungbemi IB, Oluwaseun AY, Okon EP. Fuel cell types and factors affecting them. Int J Sci Eng Res 2014;2:11–4, ISSN (Online): 2347-3878.

[33] Magro C, Paz-garcia JM, Ottosen LM, Mateus EP, Ribeiro AB. Sustainability of

        construction materials: Electrodialytic technology as a tool for mortars production. J Hazard Mater 2019;363:421–7. doi:10.1016/j.jhazmat.2018.10.010.

[34]    Zhang Z, Ottosen LM, Wu T, Jensen PE. Electro-remediation of tailings from a multi-metal sulphide mine: comparing removal efficiencies of Pb, Zn, Cu and Cd. Chem Ecol 2019;35:54–68. doi:10.1080/02757540.2018.1529173.

[35]    Tuvić T, Pašti I, Mentus S. Tungsten electrochemistry in alkaline solutions—Anodic dissolution and oxygen reduction reaction. Russ J Phys Chem A 2011;85:2399–405. doi:10.1134/S0036024411130322.

[36]    Middlehurst T. How much does a hydrogen car cost to run? Evening Stand 2017.

[37]    Baxley P, Verdugo-Peralta C, Weiss W. Rollout strategy topic team report: California 2010 Hydrogen highway network. 2005.

**Figure captions**

Figure 1. Electrodialytic cell with 3 compartments [ø = 8 cm, central and electrolyte compartments with L = 5 cm, CEM – cation exchange membrane; AEM – anion exchange membrane, An$^-$ - anions, Cat+ - cations], stirrer (only used for mine tailings suspension) connected to a proton-exchange membrane fuel cell

Figure 2. pH of the liquid phase at the central cell compartment before and after the electrodialytic experiments at 50 mA and 100 mA (error bars related to the standard deviation: n=2). BW-Briny Water; EF-Effluent; MT-Mine Tailings

Figure 3. Gas flow rate generation (mL/min) during the electrodialytic experiments with BW, EF and MTBW, at 50 mA and 100 mA (collected maximum volume 30 mL). BW-Briny Water; EF-Effluent; MT-Mine Tailings

Figure 4. Cell voltage during the electrodialytic experiments with BW, EF and MTBW, at 50 mA and 100 mA. BW-Briny Water; EF-Effluent; MT-Mine Tailings

Figure 5. Conductivity of the liquid phase at the central cell compartment before and after the electrodialytic experiments (error bars related to the standard deviation: n=2). BW-Briny Water; EF-Effluent; MT-Mine Tailings

Figure 6. Estimation of the cumulative electrical energy consumed during the electrodialytic experiments. BW-Briny Water; EF-Effluent; MT-Mine Tailings

Figure 7. Proton-exchange membrane fuel cell (PEMFC) generated open circuit voltage during the electrodialytic experiments at 100 mA. BW-Briny Water; EF-Effluent; MT-Mine Tailings

**Table legends**

Table 1. Electrodialytic experimental conditions

Table 2. Initial characterization of briny water, effluent and mine tailings

Table 3. Concentration of anions in briny water and effluent, before and after the electrodialytic experiments at 50 mA (2h) and 100 mA (1h)

Table 4. Concentration of elements in mine tailings before and after the electrodialytic experiments at 50 mA (2h) and 100 mA (1h)

Table 5. $H_2$ purity analysis by GC-TCD

## Supplementary Data

*A. Cell voltage during electrodialytic treatments: BW, EF and MTBW experiments at 50 mA or 100 mA*

| Time (min) | BW 50 mA | SD | BW 100 mA | SD | EF 50 mA | SD | EF 100 mA | SD | MTBW 50 mA | SD | MTBW 100 mA | SD |
|---|---|---|---|---|---|---|---|---|---|---|---|---|
| 0 | 16.9 | 2.4 | 30.2 | 0.4 | 19.1 | 5.1 | 31.7 | 3.8 | 18.9 | 0.2 | 34.8 | 0.7 |
| 10 | 14.8 | 2.4 | 26.2 | 0.4 | 17.6 | 6.0 | 26.4 | 5.4 | 15.8 | 0.2 | 26.6 | 1.3 |
| 20 | 13.5 | 2.3 | 23.5 | 1.0 | 16.3 | 5.9 | 24.3 | 4.0 | 14.4 | 0.1 | 25.1 | 1.9 |
| 30 | 12.7 | 2.1 | 22.2 | 0.8 | 15.6 | 5.8 | 23.1 | 3.3 | 13.3 | 0.0 | 24.4 | 2.1 |
| 40 | 12.2 | 2.1 | 21.3 | 0.5 | 15.3 | 5.9 | 22.4 | 2.8 | 13.0 | 0.0 | 24.8 | 2.3 |
| 50 | 11.9 | 2.2 | 20.7 | 0.0 | 15.2 | 6.0 | 21.7 | 2.0 | 12.8 | 0.1 | 30.2 | 2.7 |
| 60 | 11.7 | 2.3 | 20.5 | 0.4 | 15.2 | 6.0 | 21.4 | 1.8 | 12.6 | 0.1 | 41.0 | 3.4 |
| 70 | 11.7 | 2.4 | | | 15.3 | 6.1 | | | 12.4 | 0.4 | | |
| 80 | 11.8 | 2.5 | | | 15.5 | 6.0 | | | 12.3 | 0.5 | | |
| 90 | 12.2 | 2.9 | | | 15.6 | 6.1 | | | 12.3 | 0.9 | | |
| 100 | 12.6 | 3.4 | | | 15.8 | 5.9 | | | 12.5 | 1.3 | | |
| 110 | 13.3 | 3.9 | | | 15.9 | 5.6 | | | 12.9 | 2.3 | | |
| 120 | 13.7 | 4.0 | | | 16.4 | 5.2 | | | 13.2 | 2.8 | | |

*BW-Briny Water; EF-Effluent; MT-Mine Tailings; SD – Standard Deviation*

*B.1 $H_2$ volume collection during electrodialytic treatments; CEM – cation exchange membrane; AEM – anion exchange membrane*

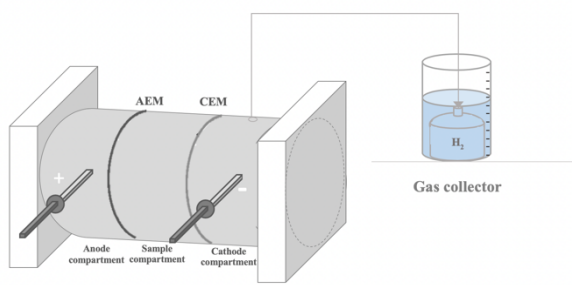

*B.2 $H_2$ generation (mL) in the electrodialytic treatments: BW, EF and MTBW experiments at 50 mA or 100 mA;*

| Time (min) | BW 50 mA | SD | BW 100 mA | SD | EF 50 mA | SD | EF 100 mA | SD | MTBW 50 mA | SD | MTBW 100 mA | SD |
|---|---|---|---|---|---|---|---|---|---|---|---|---|
| 0 | 3.0 | 4.2 | 0.0 | 0.0 | 0.0 | 0.0 | 0.0 | 0.0 | 0.0 | 0.0 | 0.0 | 0.0 |
| 10 | 9.0 | 1.4 | 11.0 | 1.4 | 10.0 | 0.0 | 17.0 | 1.4 | 11.0 | 4.2 | 15.0 | 4.2 |
| 20 | 12.0 | 0.0 | 23.0 | 1.4 | 18.0 | 2.8 | 25.5 | 0.7 | 19.0 | 5.7 | 26.0 | 0.0 |
| 30 | 17.0 | 1.4 | 30.0 | 0.0 | 25.0 | 7.1 | 30.0 | 0.0 | 25.0 | 5.7 | 30.0 | 0.0 |
| 40 | 19.0 | 1.4 | | | 26.0 | 0.0 | | | 28.0 | 6.4 | | |
| 50 | 22.0 | 2.8 | | | 30.0 | 0.0 | | | 29.0 | 4.2 | | |
| 60 | 26.0 | 2.8 | | | | | | | 30.0 | | | |
| 70 | 30.0 | 0.0 | | | | | | | | | | |

*BW-Briny Water; EF-Effluent; MT-Mine Tailings; SD – Standard Deviation*

*C. Proton-exchange membrane fuel cell generated open circuit voltage in the electrodialytic treatments: BW, EF and MTBW experiments at 100 mA;*

| Time | BW | SD | EF | SD | MTBW | SD |
|------|------|------|------|------|------|------|
| 0 | 1.35 | 0.04 | 1.36 | 0.06 | 1.41 | 0.00 |
| 10 | 1.03 | 0.71 | 0.66 | 0.67 | 1.05 | 0.00 |
| 20 | 1.03 | 0.70 | 0.60 | 0.67 | 1.05 | 0.02 |
| 30 | 1.00 | 0.68 | 0.93 | 0.17 | 1.01 | 0.04 |
| 40 | 0.98 | 0.69 | 0.92 | 0.15 | 0.99 | 0.04 |
| 50 | 0.97 | 0.63 | 0.94 | 0.10 | 0.99 | 0.01 |
| 60 | 0.96 | 0.64 | 0.94 | 0.08 | 0.98 | 0.00 |

*BW-Briny Water; EF-Effluent; MT-Mine Tailings; SD – Standard Deviation*

*D. Electrodialytic $H_2$ flow - ANOVA with 95% of confidence level*

| $y = a + b*x$ | Matrices | | | | | |
|---|---|---|---|---|---|---|
| | BW 50 mA | BW 100 mA | EF 50 mA | EF 100 mA | MTBW 50 mA | MTBW 100 mA |
| Residual SoS | 54.74 | 6.50 | 68.18 | 56.11 | 173.53 | 40.36 |
| Pearson's r | 0.99 | 1.00 | 0.99 | 0.98 | 0.98 | 0.99 |
| Adj. R-Square | 0.98 | 0.99 | 0.97 | 0.96 | 0.94 | 0.97 |
| **Slope (flow)** | **0.45$^a$** | **1.05$^b$** | **0.68$^A$** | **1.13$^{Bc}$** | **0.62$^d$** | **1.12$^{Ce}$** |
| Slope standard Error | 0.02 | 0.04 | 0.05 | 0.12 | 0.06 | 0.10 |

*Statistical analysis: Multiple comparisons were statistically performed at p<0.05 (95% confidence interval); data with lower case letters is statistically significantly different to the ones with the same capital letter.*

*BW-Briny Water; EF-Effluent; MT-Mine Tailings*

*E. Comparing electrodialytic experiments statistics on $H_2$ flow - ANOVA with 95% of confidence level*

| **ANOVA summary ED 50mA** | |
|---|---|
| F | 13.14 |
| P value | 0.0328 |
| P value summary | * |
| Significant diff. among means *(P < 0.05)?* | **Yes** |
| R square | 0.8975 |
| *Combinations Significantly different* | |
| BW 50 mA | EF 50 mA |

| **ANOVA summary ED 100mA** | |
|---|---|
| F | 0.4385 |
| P value | 0.6807 |
| P value summary | ns |
| Significant diff. among means *(P < 0.05)?* | **No** |
| R square | 0.2262 |

*F. pH and conductivity data and statistics: ANOVA with 95% of confidence level*

|  | Initial | | | ED Final | |
|---|---|---|---|---|---|
|  | pH | Conductivity (mS/cm) |  | pH | Conductivity (mS/cm) |
| Electrolyte | 6.46 ± 0.55[a] | 0.90 ± 0.06 | BW 50 mA Anolyte | 2.02 ± 0.21 | 4.40 ± 0.28 |
| BW | 6.89 ± 0.08 | 1.81 ± 0.11 | BW 50 mA Catholyte | 12.29 ± 0.14 | 3.00 ± 0.71 |
| EF | 7.67 ± 0.16[A] | 2.41 ± 0.69 | BW 50 mA Central | 6.08 ± 0.59 | 1.15 ± 0.33 |
| MTBW | 7.03 ± 0.06 | 1.82 ± 0.54 | BW 100 mA Anolyte | 2.18 ± 0.3 | 2.95 ± 0.49 |
|  |  |  | BW 100 mA Catholyte | 12.13 ± 0.20 | 1.23 ± 1.65 |
|  |  |  | BW 100 mA Central | 3.19 ± 0.04 | 1.24 ± 0.45 |
|  |  |  | EF 50 mA Anolyte | 2.20 ± 0.02 | 2.70 ± 0.00 |
|  |  |  | EF 50 mA Catholyte | 12.21 ± 0.01 | 2.45 ± 0.21 |
|  |  |  | EF 50 mA Central | 4.54 ± 2.55 | 1.80 ± 0.86 |
|  |  |  | EF 100 mA Anolyte | 2.21 ± 0.04 | 2.85 ± 1.06 |
|  |  |  | EF 100 mA Catholyte | 12.11 ± 0.01 | 1.94 ± 0.23 |
|  |  |  | EF 100 mA Central | 2.85 ± 0.20 | 1.49 ± 0.15 |
|  |  |  | MTBW 50 mA Anolyte | 2.02 ± 0.11 | 3.25 ± 0.21 |
|  |  |  | MTBW 50 mA Catholyte | 12.30 ± 0.01 | 2.40 ± 0.14 |
|  |  |  | MTBW 50 mA Central | 5.91 ± 0.45 | 1.41 ± 0.49 |
|  |  |  | MTBW 100 mA Anolyte | 1.99 ± 0.04 | 2.55 ± 0.07 |
|  |  |  | MTBW 100 mA Catholyte | 12.21 ± 0.25 | 2.08 ± 0.46 |
|  |  |  | MTBW 100 mA Central | 4.42 ± 1.33 | 0.24 ± 0.16 |

*Statistical analysis: Multiple comparisons were statistically performed at p<0.05 (95% confidence interval); data with lower case letters is statically significantly different to the ones with the same capital letter.*
*BW-Briny Water; EF-Effluent; MT-Mine Tailings*